\documentclass[12pt]{article}
\begin{document}

\title{\bf  Collapsing Plane Symmetric Source with Heat Flux and Conformal Flatness}

\author{ G.
Abbas$^{(a)}$  \thanks{ghulamabbas@ciitsahiwal.edu.pk}, Zahid
Ahmad$^{(b)}$ \thanks{zahidahmad@ciit.net.pk} and Hassan Shah
$^{(b)}$ \thanks{hassanshah13@yahoo.com}\\\\
$^{(a)}$  Department of Mathematics, COMSATS\\
Institute of Information Technology,  Sahiwal-57000, Pakistan.\\
$^{(b)}$ Department of Mathematics, COMSATS\\
Institute of Information Technology, Abbottabad-22060, Pakistan.}

%\author{G. Abbas$^{(a)}$ \thanks{ghulamabbas@ciitsahiwal.edu.pk}. Zahid Ahmad$^{(b)}$ \thanks{zahidahmad@ciit.net.pk} and Hassan Shah $^{(b)}$
% \thanks{hassanshah13@yahoo.com}  \\
%%EndAName
%
%%$^a$\textit{{Department of Mathematics, COMSATS Institute of
%%\\Information
%%Technology Sahiwal-57000, Pakistan}}
%
%\textit{\ {\small Department of Mathematics, }} \\
%\textit{{\small COMSATS Institute of Information Technology,}} \\
%\textit{{\small University Road, Post Code 22060, Abbottabad, KPK,
%Pakistan }}}
\date{}
\maketitle

\begin{abstract}
This paper deals with the study of collapsing plane symmetric source
in the presence of heat flux. For this purpose, we have calculated
the Einstein field equations as well as Weyl tensor components. The
conditions for the conformal flatness have been determined. The
interior source has been matched smoothly with the exterior geometry
in single null coordinate. It has been found the pressure is
balanced with the out going heat flux and the continuity of the
masses in two regions has been noted. A simple new model of collapse
has been proposed which satisfies flatness condition, also we have
discussed the physical properties of the model. For our model, we
have
calculated the temperature profile by using the approximation scheme.\\

\end{abstract}

 PACS numbers:
$04.20$.Dw, $04.25$.Nx, $04.30$.Db\\
 {\bf Keywords }: Gravitational
Collapse, Matching Conditions, Heat Transport Equations.

\section{Introduction}

The concept of gravity in terms of spacetime curvature was
introduced for the first time by Einstein. According to him,
curvature in the spacetime is produced due to the presence of matter
in which it exists. This curvature is taken as a source of
gravitational field. He formulated a set of equations known as
Einstein field equations. The solution of these equations have
spacetime singularities. These singularities are formed by
gravitational collapse. Gravitational collapse is one of the
important problem in general relativity.

In astronomy a gravitational collapse is a phenomenon in which
a massive star dozens of times larger than the solar mass
contracts under the influence of its own gravity.
It occurs when the internal nuclear fuel fails to supply
sufficiently high pressure to counterbalance the gravity.
There are two types of spacetime singularities,
one is a black hole and the other is a naked.
The singularity covered by an event horizon is called
black hole and an uncovered singularity is called naked.
Event horizon is the boundary of the black hole.

Many researchers investigated the solutions of Einstein field
equations with heat flux (Schafer and Goenner 2000, Chan et al.2003,
Herrera and Santos 2003). Ghosh and Deshker (2007) investigated
exact non spherical radiating collapse using junction conditions.
Ahmad et al. (2013) investigated the gravitational collapse with
heat flux and gravitational waves.

Initially Bronnikov and Kovalchuk (1983) studied cylindrical
symmetry and found some exact solutions. The collapse of dust
spheroid was studied numerically by Shapiro and Teukolsky (1991)
using cylindrical spacetime. They concluded that the end product of
collapse was either a black hole or a naked depending on the
spheroid compactness. Barrabes et al.(1991) investigated the exact
solution for collapsing convex shell. They concluded that in some
cases they found the absence of apparent horizons. Sharif and Ahmad
(2007) studied high-speed cylindrical collapse of two perfect
fluids.

Gutti et al.(2012) investigated the gravitational collapse of an
infinite cylindrical distribution of timelike dust, using the
matching conditions. The part of anisotropy, radial heat flux and
electric charge over the dynamics of non adiabatic cylindrical
collapse by using the coupled equation was studied by Sharif and
Abbas (2011). Abbas and his collaborators (abbas 2014a, Abbas 2014b
Mahmood et al.2015) investigated the expanding and collapsing
solutions for charged plane and cylindrical stars. The shearfree
cylindrical collapse was studied by Di Prisco et al.(2009). Herrera
et al. (2004) studied spherical shearfree radiating collapse and
conformal flatness they developed a relation between dissipation and
density inhomogeneity. Also, Herrera and Santos (1997) explored the
properties of anisotropic self-gravitating spheres and discussed
their stability using the perturbation method. Herrera and his
collaborators (Herrera et al.2008a, Herrera et al. 2008b, Herrera at
al. 1989, Herrera et al. 2009a, Herrera et al. 2010, Herrera et al.
2012) have discussed the stability and applications of anisotropic
solutions to stellar collapse.

In this paper, the work done by Herrera et al.(2004) is extended by
taking the plane symmetric gravitational with heat flux. The plan of
the paper is as follows:  The matter source and field equations are
presented in section \textbf{{2}}. In section $3$, we provide the
formulation of solutions by taking into account conformal flatness.
In section $4$, we construct a simple dissipative model and studied
the temperature profile of the proposed model. We conclude the
results in the last section.

\section{Interior Matter Distribution and Field Equations}
Assume that a given four dimensional plane symmetric spacetime is
divided by a three dimensional hypersuface $\Sigma$
into two regions interior and exterior denoted by $V^{-}$and $%
V^{+}\ $respectively. We consider a particular plane symmetric
spacetime in interior region given by
\begin{equation}
ds_{-}^{2}=-A^{2}(t, z)dt^{2}+B^{2}(t, z)[dx^{2}+dy^{2}+dz ^{2}].
\label{1}
\end{equation}%
Here the coordinates are labeled as follows: $x^{0}=t, x^{1}=x,
x^{2}=y$ and $x^{3}=z$. We assume that the energy momentum tensor in
the interior region is given by
\begin{equation}
T^{-}_{\alpha \beta }=\left( \mu +P \right) w_{\alpha
}w_{\beta }+Pg_{\alpha \beta }+w_{\alpha }q_{\beta }+q_{\alpha
}w_{\beta }, \label{2}
\end{equation}
where $\mu \ $ is the energy density, $P$ the
pressure,$w^{\alpha }\ $ the four velocity of fluid and $q^{\alpha }$
the heat flux satisfying $q_{\alpha }w^{\alpha }=0 $.
For interior spacetime (\ref{1}) in comoving coordinates, we have
\begin{eqnarray}
w^{\alpha }=\frac{1}{A}\delta ^{\alpha}_{0 },\label{3}\\
q^{\alpha }=q\delta_{3}^{\alpha }, \label{4}
\end{eqnarray}
where $q$ depend on $t$ and $z$.

In this case the rate the rate of expansion scalar
$\Theta=w^{\alpha}_{;\alpha}$  is given by
\begin{equation}
\Theta=3\frac{B_{,t}}{AB}, \label{5}
\end{equation}

 Weyl tensor $C_{\alpha\beta\gamma\delta}$ has the following non-zero components for metric (\ref{1})
\begin{eqnarray}
C_{1212}=\frac{B^{2}}{3}\left[2\frac{B_{,z}}{B}\left(\frac{A_{,z}}{A}
-\frac{B_{,z}}{B}\right)-\left(\frac{A_{,zz}}{A}-\frac{B_{,zz}}{B}\right)\right], \label{6}\\
C_{1212}=2\left(\frac{B}{A}\right)^{2}C_{0101}=-\left(\frac{B}{A}\right)^{2}C_{0303}=-2C_{1313}. \label{7}
\end{eqnarray}
For the interior metric (\ref{1}) the non vanishing components of Einstein field equations
are given by

\begin{eqnarray}
G_{00}^{-}
&=&\left(\frac{A}{B}\right)^{2}\left(-2\frac{B_{,zz}}{B}+\left(\frac{B_{,z}}{B}\right)^{2}\right)+3\left(\frac{B_{,t}}{B}\right)^{2}
=\kappa \mu A^{2},  \label{8} \\
G_{11}^{-} &=&-\left(\frac{B}{A}\right)^{2}\left(2\frac{B_{,tt}}{B}-2\frac{A_{,t}B_{,t}}{AB}%
+\left(\frac{B_{,t}}{B}\right)^{2}\right)
\nonumber\\&&+\frac{A_{,zz}}{A}+\frac{B_{,zz}}{B}-\left(\frac{B_{,z}}{B}\right)^{2}%
=\kappa PB^{2},  \label{9} \\
G_{22}^{-} &=&G_{11}^{-},\nonumber \\
G_{33}^{-} &=&-\left(\frac{B}{A}\right)^{2}\left(2\frac{B_{,tt}}{B}-2\frac{A_{,t}B{,t}}{AB}+\left(\frac{B_{,t}}{%
B}\right)^{2}\right)
\nonumber\\&&+2\frac{A_{,z}B_{,z}}{AB}+\left(\frac{B_{,z}}{B}\right)^{2}=\kappa PB^{2},  \label{10}\\
G_{03}^{-} &=&-2\left(\frac{B_{,t}}{AB}\right)_{,z}A%
=-\kappa qAB^{2}.  \label{11}
\end{eqnarray}
Using Eqs.(\ref{11}) and (\ref{5}), we get
\begin{equation}
\kappa qB^{2}=\frac{2}{3}\Theta_{,z}, \label{12}
\end{equation}
This indicates that for $q>0$ heat flux is directed outward,
implying that $\Theta_{,z}>0$. For $q=0$,  Eq. (\ref{12}) implies
that $\Theta_{,z}=0$ which shows that collapse is homogeneous
(Herrera and Santos 2003). The Taub's mass function (Zannias 1990)
for the plane symmetric spacetime is given by
\begin{equation}
m(t, z)=\frac{(g_{11})^{\frac{3}{2}}}{2}R_{12}^{12}=\frac{B}{2}\left[\frac{B_{,t}^{2}}{A^{2}}-\frac{B_{,z}^{2}}{B^{2}}\right], \label{13}
\end{equation}
Differentiation of $m(t,z)$ with respect to $z$ and $t$ along with field equations (\ref{8})-(\ref{11}) gives
\begin{eqnarray}
m'(t, z)=\frac{\kappa}{2}\left[B^{2}B_{,z}\mu+qB^{4}\frac{B_{,t}}{A}\right], \label{14}\\
\dot{m}(t, z)=-\frac{\kappa}{2}\left[PB^{2}B_{,t}+qB^{2}B_{,z}A\right]. \label{15}
\end{eqnarray}
The above equations show that the heat flux may effect the gradient
and time derivative of $m(z, t)$, and the dissipation reduces the
amount of matter, hence the rate of collapse slows down probably.

 The scalar of the Weyl tensor in terms of Kretchsmann
scalar
${\tilde{R}}^{2}=R^{\alpha\beta\gamma\delta}R_{\alpha\beta\gamma\delta}$,
the Ricci tensor $R_{\alpha\beta}$ and Ricci scalar $R$ is defined
as
\begin{equation}
C^{2}=\tilde{R}-2R^{\alpha\beta}R_{\alpha\beta}+\frac{1}{3}R^{2}.
\label{16}
\end{equation}
Using Eqs.(\ref{6}), (\ref{8}) and (\ref{13}) in Eq. (\ref{16}), we
obtain Weyl tensor in terms of pure gravitational mass given by
\begin{equation}
C^{2}=48\left[\frac{m}{B^{3}}-\frac{\kappa\mu}{6}\right]^{2}=48\frac{m^{2}_{c}}{B^{6}}, \label{17}
\end{equation}
where $m_{c}= m-\frac{\kappa}{6}\mu B^{3}$.

\section{Conformally flat solution}

In this section, we impose conformal flatness condition to metric
(\ref{1}), this requirement implies that all the components of the
Weyl tensor must be equal to zero. Hence, from  Eqs.(\ref{6}) and
(\ref{7}), we see that if $C_{1212}=0$, this condition is justified
and we have
\begin{equation}
\left(\frac{A_{,zz}}{A}-\frac{B_{,zz}}{B}\right)-2\frac{B_{,z}}{B}\left(\frac{A_{,z}}{A}
-\frac{B_{,z}}{B}\right)=0. \label{19}
\end{equation}
Integrating above equation and after reparamatrizing $t$, we get
\begin{equation}
A=\left[C_{1}(t)z+1\right]B. \label{20}
\end{equation}
where $C_{1}$ is arbitrary function of $t$. Also, from the isotropy
of pressure $G_{11}=G_{33}$ with Eq.(\ref{20}), gives
\begin{equation}
\frac{B_{,zz}}{B_{,z}}-2\frac{B_{,z}}{B}=0, \label{21}
\end{equation}
integration of Eq.(\ref{21}) yields
\begin{equation}
B=\frac{1}{C_{2}(t)z+C_{3}(t)}, \label{22}
\end{equation}
where $C_{2}$ and $C_{3}$ are the constant of integration depends on
$t$. Stephani (1967) and Kramer et al.(1981) investigated all
conformally flat solutions with $q=0$.

Conformal flatness imposes $\emph{C}=0$ and put $m_{c}=0$ in
Eq.(\ref{17}), we get
\begin{equation}
m=\frac{\kappa}{6}\mu B^{3}. \label{23}
\end{equation}
Using Eqs. (\ref{14}) and (\ref{23}) we get
\begin{equation}
\mu'=qB^{2}\Theta, \label{24}
\end{equation}
which describes that for $q>0$ and $\Theta<0$ then $\mu'<0$ implying that the density reduces with
increasing $z$, while from (\ref{5}) and (\ref{12}) we get
\begin{equation}
\kappa\mu'=\frac{1}{3}(\Theta^{2})_{,z}, \label{25}
\end{equation}
integrating (\ref{25}) we get
\begin{equation}
\kappa\mu=\frac{\Theta^{2}}{3}+g(t), \label{26}
\end{equation}
where $g$ depend on $t$ only.

Putting solution (\ref{20}) and (\ref{22}) into (\ref{8}), (\ref{9}) and (\ref{11}) it follows
\begin{eqnarray}
\kappa\mu &=&3\left(\frac{\dot{C}_{2}z+\dot{C}_{3}}{C_{1}z+1}\right)^{2}-4C_{2}^{2}, \label{27}\\
\kappa P &=&\frac{1}{C_{1}z+1}[2(\ddot{C}_{2}z+\ddot{C}_{3})(C_{2}z+C_{3})-3(\dot{C}_{2}z+\dot{C}_{3})^{2}
\nonumber\\&&-\frac{2\dot{C}_{1}}{C_{1}z+1}(\dot{C}_{2}z+\dot{C}_{3})(C_{2}z+C_{3})z]
\nonumber\\&&+\frac{1}{C_{1}z+1}
[C_{1}C_{2}^{2}z+3C_{2}^{2}-2C_{1}C_{2}C_{3}], \label{28}\\
\kappa q &=&2({C}_{1}\dot{C}_{3}-\dot{C}_{2})\left(\frac{C_{2}z+C_{3}}{C_{1}z+1}\right)^{2}.\label{29}
\end{eqnarray}
The expansion of the fluid given by (\ref{5}) along with Eqs.(\ref{20}), (\ref{22}) and (\ref{29}), yields
\begin{equation}
\Theta=-3\frac{\dot{C}_{2}z+\dot{C}_{3}}{C_{1}z+1}=-3\left[\dot{C}_{3}
-\frac{\kappa qz}{2}\frac{{C_{1}z+1}}{(C_{2}z+{C}_{3})^{2}}\right]. \label{30}
\end{equation}
The above equation show that for $q=0$ the contraction is
homogeneous, however for $q\neq0$, collapse is inhomogeneous, which
has already been concluded in (\ref{12}). From Eqs. (\ref{27}) and
(\ref{26}), we see that $g(t)=-4C_{2}^{2}$.  Here, we matched the
interior spacetime with the exterior spacetime that is described by
plane symmetric spacetime in single null coordinate given by
\begin{equation}
ds_{+}^{2}=\frac{2m(\nu)}{z}d\nu^{2}-2d\nu dz+z^{2}(dy^{2}+dz ^{2}),
\label{31}
\end{equation}%
from Eqs.(\ref{1}) and (\ref{31}) on hypersurface $\Sigma$ along with field equations (\ref{8})-(\ref{11})
and the mass function (\ref{13}), we get
\begin{eqnarray}
B_{\Sigma} &=&z_{\Sigma}, \label{32}\\
P_{\Sigma} &=&(qB)_{\Sigma}, \label{33}\\
m(\nu) &=&\frac{B}{2}\left[\frac{B_{,t}^{2}}{A^{2}}-\frac{B_{,z}^{2}}{B^{2}}\right]_{\Sigma}. \label{34}
\end{eqnarray}
Using Eqs. (\ref{28})-(\ref{29}) and (\ref{33}) we obtain
\begin{eqnarray}
&&\Big[\ddot{C}_{2}z+\ddot{C}_{3}-\frac{3}{2}\frac{(\dot{C}_{2}z+\dot{C}_{3})^{2}}{C_{2}z+C_{3}}
-(C_{1}\dot{C}_{3}-\dot{C}_{2})
\nonumber\\&&-\frac{2\dot{C}_{1}z(\dot{C}_{2}z+\dot{C}_{3})}{C_{1}z+1
}+\frac{(C_{1}z+1)^{2}}{2(C_{2}z+C_{3})}
\nonumber\\&&(C_{1}C_{2}^{2}z+3C_{2}^{2}-2C_{1}C_{2}C_{3})\Big]=0
. \label{35}
\end{eqnarray}

\section{A simple model}
A simple approximation solution satisfying the junction condition (\ref{35}) for the
functions $C_{1}(t), C_{2}(t)$ and $C_{3}(t)$ is
\begin{equation}
C_{1}=\epsilon c_{1}(t),\ C_{2}(t)=0,\ C_{3}(t)=\frac{a}{t^{2}}, \label{36}
\end{equation}%
where $0<\epsilon<<1$,\ $0\leq t\leq -\infty$ and $a>0$ a constant proportional to
the total mass in the plane.
Using (\ref{36}) and (\ref{35}) we get upto order of $\epsilon$,
\begin{equation}
\dot{c}_{1}+\frac{c_{1}}{z_{\Sigma}}\approx0, \label{37}
\end{equation}
integrating (\ref{37}) we get,
\begin{equation}
{c}_{1}(t)\approx c_{1}(0)exp[-\frac{t}{z_{\Sigma}}]. \label{38}
\end{equation}
Putting the solution (\ref{36}) into (\ref{27})-(\ref{29}), we get
\begin{eqnarray}
\kappa\mu &\approx&\frac{12a^{2}}{t^{6}}(1-2\epsilon c_{1}z), \label{39}\\
\kappa P &\approx&-\frac{4\epsilon c_{1}a^{2}}{t^{5}}\frac{z}{z_{\Sigma}}, \label{40}\\
\kappa q &\approx&-\frac{4\epsilon c_{1}a^{3}}{t^{7}},\label{41}
\end{eqnarray}
 above conditions hold for a realistic process in stellar
 evaluation. We would like to mention that for
$\epsilon\neq0$ the value of $t$ is constrained by some physical
situation. Thus, if one requires that the central pressure remain
less than the value of the central energy density, then there exists
the inequality
\begin{equation}
\frac{3}{t_{2}}>\epsilon c_{1}.\label{42}
\end{equation}

From Eq.(\ref{39}) and Eq.(\ref{40}), we see that the energy density
and pressure decrease in the outer region due to dissipation. It can
be noted from (\ref{41}), that there is increase in the heat flow
outwardly.

From (\ref{13}) with (\ref{20}) and (\ref{22}) we get
\begin{equation}
m(t, z)\approx\frac{2}{a}(1-2\epsilon c_{1}z), \label{43}
\end{equation}
which indicates that mass of the gravitating system is being
decreased due to heat flux.  Now the expansion scalar (\ref{36})
becomes
\begin{equation}
\Theta\approx\frac{6a}{t^{3}}(1-\epsilon c_{1}z).\label{44}
\end{equation}
The above equation shows that dissipation causes to decrease the
rate of collapse. The adiabatic index in the presence of heat flux
is defined as follows
\begin{equation}
\Gamma_{eff}=\frac{dlnP}{dln\mu}=\left(\frac{\dot{P}}{P}\right)
\left(\frac{\mu}{\dot{\mu}}\right),\label{45}
\end{equation}
It gives the dynamical instability at a particular time, now
computing (\ref{45}) at  $z=0$ and $z=z_{\Sigma}$ (using
Eq.(\ref{36}) and (\ref{38})) upto order of $\epsilon$, we get
\begin{eqnarray}
\Gamma_{z=0}\approx\frac{t}{6z_{\Sigma}}+\frac{5}{6}, \label{46}\\
\Gamma_{z=z_{\Sigma}}\approx\frac{t}{6z_{\Sigma}}+\frac{5}{6}. \label{47}
\end{eqnarray}
The above equations imply that the collapse is homogeneous at the
center and on the hypersurface $\Sigma$.

\subsection{Calculation of Temperature}
This section deals with the calculation of temperature profile $T(z,
t)$ for our model. The heat transport equation in Maxwell-Cattaneo
theory (Herrera et al.2004) is
\begin{equation}
\tau h^{\alpha\beta}w^{\gamma}q_{\beta;\gamma}+q^{\alpha}=-Kh^{\alpha\beta}(T_{,\beta}+a_{\beta}T),\label{48}
\end{equation}
where the quantities $\tau, K$ and
$h^{\alpha\beta}=g^{\alpha\beta}+w^{\alpha}w^{\beta}$ are the time
relaxation, thermal conductivity and the projection tensor
respectively. Considering Eqs. (\ref{1})-(\ref{4}) then (\ref{48})
becomes
\begin{equation}
\tau B(Bq)_{,t}+qAB^{2}=-K(TA)_{,z},\label{49}
\end{equation}
Applying Eqs.(\ref{20}), (\ref{22}) and (\ref{29}) into (\ref{49})
and taking $C_{2}=0$, we obtain upto $O(\epsilon)$
\begin{equation}
\tau (C_{1}C_{3}\dot{C}_{3})_{,t}+C_{1}\dot{C}_{3}=-\frac{\kappa K}{2}[T(c_{1}z+1)]_{,z}.\label{50}
\end{equation}
Now if there is no dissipation we have $(C_{1}=\epsilon c_{1}=0)$
from (\ref{50}) it follows that $T=T_{0}(t)$ this shows that for
non-dissipative systems, the temperature is homogenous through out
the system. Thus for dissipative case $C_{1}\neq0$, we have the
following form of temperature profile
\begin{equation}
T=T_{0}(t)+\epsilon T_{\epsilon}(z, t).\label{51}
\end{equation}
substituting (\ref{51}) into (\ref{50}) we get upto $O(\epsilon)$
\begin{equation}
T\approx T_{c}(t)+\epsilon c_{1}\left(\frac{4a}{\kappa Kt^{3}}-T_{0}(t)\right)z
-\frac{4\tau\epsilon c_{1}a^{2}}{\kappa Kt^{5}}\left(\frac{5}{t}+\frac{1}{z_{\Sigma}}\right)z.\label{52}
\end{equation}
Here, it is assumed that $K>0$ and $T_{c}$ is the central
temperature. The dissipation can decrease the temperature at the
loss of energy, as it is evident from second term of term of above
equation while last term predicts the support of relaxational
effects.

\section{Conclusion}

It is well known that gravitational collapse of a stellar object is
extremely dissipative process. Therefore, it becomes important to
study the effects of dissipation during the collapse of a radiating
star. During the dynamical process of gravitational collapse of a
non-adiabatic star, the enough amount of energy losses in the form
of outward heat flow and radiations. Herrera et al.(2009a) derived
 the dynamical equations by including the dissipation term in the source in the form of heat
flow, radiation, shear and bulk viscosity and then coupled the field
equations with the causal transport equation. The inertia due to
heat flux and its effect during the dynamics of dissipative collapse
with outgoing radiations was studied by Herrera (2006).

In this paper, we investigate the plane gravitational collapse of
radiating fluid with conformal flatness condition. For this purpose,
we have explored some the effects of dissipation on the dynamics of
self-gravitating body. By determining the conformal flatness
condition of the collapsing star, we have found that the system of
equations is exactly solvable and there appears three arbitrary
integrating functions which depend on $t$. This implies that
conformal flatness conditions to leads to the homogeneous solutions.
Further, the interior solution has been matched smoothly with the
plane symmetric metric which is defined in single null coordinate.
We would like to mention that exterior geometry emits the radiations
in the form of outgoing radiations. We constructed a simple model
which satisfies the matching conditions.

We have calculated the adiabatic index which shows that gravitating
system is instable at particular instant of time. The perturbation
analysis of the adiabatic index implies that the collapse is
homogeneous at the center as well as on the hypersurface $\Sigma$.
Using the Maxwell-Cattaneo heat transport equation, we have
calculated the temperature profile of the radiating star. The
importance of the proposed model lies in the fact that it presents
the influence of relaxational effects on the temperature profile in
a sophisticated way and thereby on the evolution of the system. It
is concluded that the inhomogeneities in the energy density are
directly related to dissipation, even though the underlying
spacetime is conformally flat.

\section{Conflict of Interest} The authors declare that they have no
conflict of interest.

\end{document}